\documentclass[twocolumn, 5p, 12pt]{elsarticle}

\pdfoutput=1

\usepackage{graphicx}
\usepackage{subfigure}
\usepackage[dvipdfx,cmyk]{xcolor}
\usepackage{amssymb}

\biboptions{sort&compress}

\journal{Physics Letters A}

\usepackage[utf8]{inputenc}
\usepackage[T1]{fontenc}
\usepackage{graphicx}
\usepackage{float}
\usepackage{amssymb}
\usepackage{amsmath}
\usepackage{hyperref}
\usepackage{subfigure}
\usepackage{color}
\usepackage{multirow}
\usepackage{relsize}
\usepackage{url}

\definecolor{amethyst}{rgb}{0.6, 0.4, 0.8}
\definecolor{blue-violet}{rgb}{0.54, 0.17, 0.89}
\definecolor{blue(pigment)}{rgb}{0.2, 0.2, 0.6}
\definecolor{byzantium}{rgb}{0.44, 0.16, 0.39}
\definecolor{coolblack}{rgb}{0.0, 0.18, 0.39}
\definecolor{english}{rgb}{0.0, 0.5, 0.0}
\definecolor{armygreen}{rgb}{0.29, 0.33, 0.13}
\definecolor{red}{rgb}{1,0,0} 
\definecolor{yellow}{rgb}{0.8, 0.8, 0} 
\definecolor{blue}{rgb}{.05,.05,.9} 
\definecolor{purple}{rgb}{0.5,0,0.5} 
\definecolor{green}{rgb}{0,0.5,0} 
\definecolor{orange}{rgb}{.8,0.5,0} 
\definecolor{pink}{rgb}{.8,0,0.8} 

\newcommand\GR{{G_{\rm R}}}
\newcommand\Grr{{G_{rr}}}
\newcommand\Gss{{G_{ss}}}
\newcommand\Grs{{G_{rs}}}
\newcommand\Gsr{{G_{sr}}}
\newcommand\Gqr{{G_{\rm qr}}}
\newcommand\Gpr{{G_{\rm pr}}}
\newcommand\Gqrd{{G_{\rm qrd}}}
\newcommand\Gqrnd{{G_{\rm qrnd}}}
\newcommand\Wqr{{W_{\rm qr}}}
\newcommand\Wpr{{W_{\rm pr}}}
\newcommand\Wrr{{W_{rr}}}

\begin{document}

\begin{frontmatter}
\title{Multi-cultural Wikipedia mining of geopolitics interactions leveraging reduced Google matrix analysis}

\author[klaus]{Klaus M.~Frahm}
\ead{frahm@irsamc.ups-tlse.fr}

\address[klaus]{Laboratoire de Physique Th\'{e}orique du CNRS, IRSAMC, Universit\'{e} de Toulouse, UPS, F-31062 Toulouse, France}

\author[katia]{Samer El Zant}
\ead{samer.elzant@enseeiht.fr}


\author[katia]{Katia Jaffr\`es-Runser}
\ead{kjr@enseeiht.fr}

\address[katia]{Institut de Recherche en Informatique de Toulouse, Universit\'e de Toulouse, INPT, 31061 Toulouse, France}

\author[dima]{D.~Shepelyansky}
\ead{dima@irsamc.ups-tlse.fr}

\address[dima]{Laboratoire de Physique Th\'{e}orique du CNRS, IRSAMC, Universit\'{e} de Toulouse, UPS, F-31062 Toulouse, France}

\begin{abstract}
Geopolitics focuses on political power in relation to geographic space. Interactions among world countries have been widely studied at various scales, observing economic exchanges, world history or international politics among others.  
This work exhibits the potential of Wikipedia mining for such studies. Indeed, Wikipedia stores valuable fine-grained dependencies among countries by linking webpages together for diverse types of interactions (not only related to economical, political or historical facts). We mine herein the Wikipedia networks of several language editions using the recently proposed method of reduced Google matrix analysis. This approach allows to establish direct and hidden links between a subset of nodes that belong to a much larger directed network. Our study concentrates on 40 major countries chosen worldwide. 
Our aim is to offer a multicultural perspective on their interactions by comparing networks extracted from five different Wikipedia language editions, emphasizing English, Russian and Arabic ones. We demonstrate that this approach allows to recover meaningful direct and hidden links among the 40 countries of interest.
\end{abstract}




\end{frontmatter}

\section{Introduction}\label{sec:intro}

Political and economic interactions between regions of the world have always been of utmost interest to measure and predict their relative influence. Such studies belong to the field of geopolitics that focuses on political power in relation to geographic space. Interactions among world countries have been widely studied at various scales (worldwide, continental or regional) using different types of information. Studies are driven by observing economic exchanges, social changes, history, international politics and diplomacy among others \cite{jones, dittmer}. 
The major finding of this paper is to show that meaningful worldwide interactions can be automatically extracted from the global and free online Encyclopaedia Wikipedia \cite{wikiorg} for a given set of countries. All information gathered in this collaborative knowledge base can be leveraged to provide a picture of countries relationships, fostering a new framework for thorough geopolitics studies.

Wikipedia has become the largest open source of knowledge being close to Encyclopaedia Britanica \cite{britanica} by the accuracy of its scientific entries \cite{giles} and overcoming the latter by the enormous quantity of available information. A detailed analysis of strong and weak features of Wikipedia is given at \cite{reagle,finn}. Wikipedia articles make citations to each other, providing a direct relationship between webpages and topics. As such, Wikipedia generates a larger directed network of article titles with a rather clear meaning. For these reasons, it is interesting to apply algorithms developed for search engines of World Wide Web (WWW) such as the PageRank algorithm \cite{brin}(see also \cite{meyer}), to analyze the ranking properties and relations between Wikipedia articles. For various language editions of Wikipedia it was shown that the PageRank vector produces a reliable ranking of historical figures over 35 centuries of human history \cite{wikizzs,wikievol,eomwiki9,eomwiki24,rmp2015} and a solid Wikipedia ranking of world universities (WRWU) \cite{wikizzs,lageswiki}. It has been shown that the Wikipedia ranking of historical figures is in a good agreement with the well-known Hart ranking \cite{hart}, while the WRWU is in a good agreement with the Shanghai Academic ranking of world universities \cite{shanghai}. 

\begin{figure}[h]
\begin{center}
\includegraphics[width=0.36\textwidth]{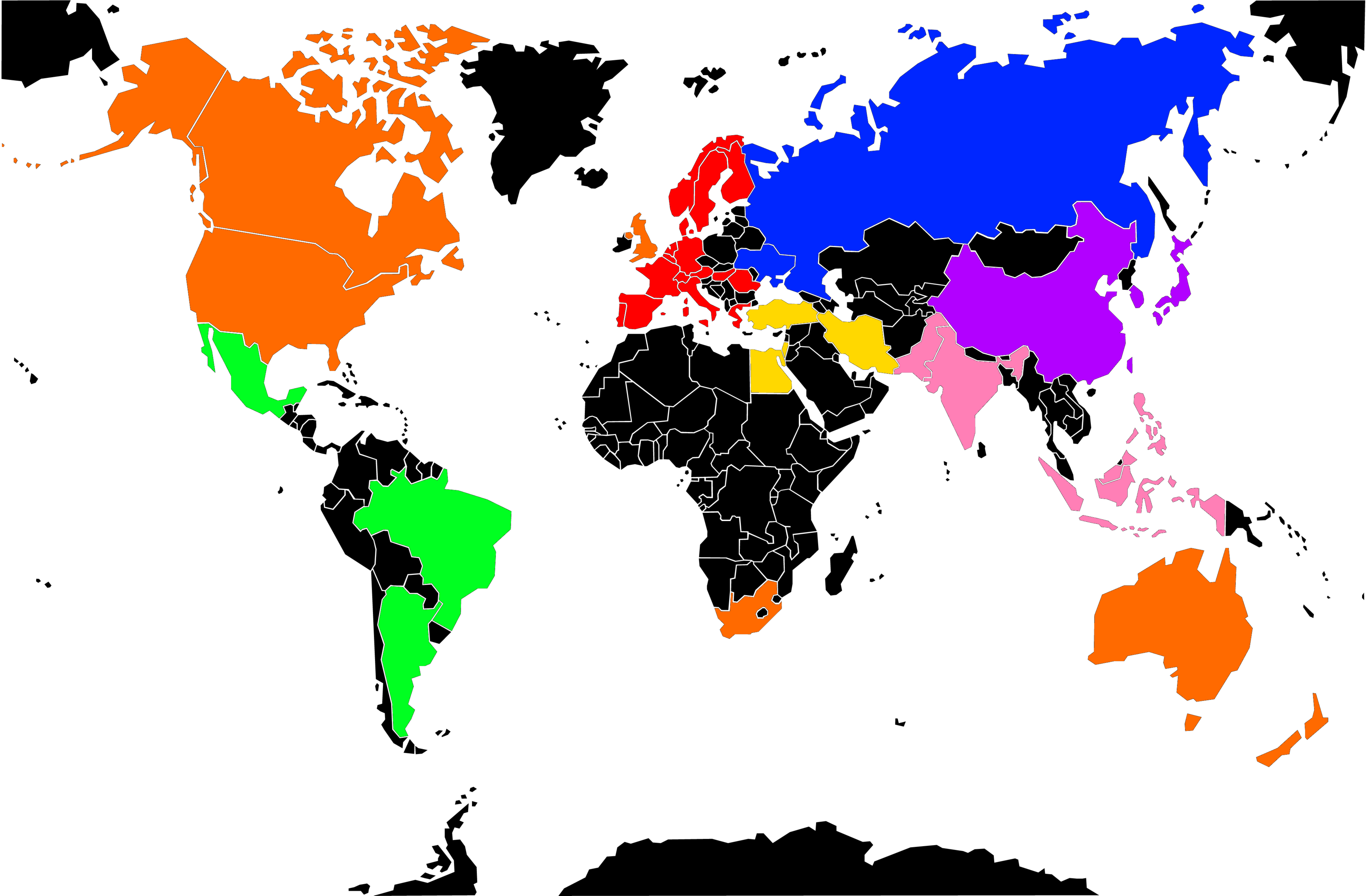}
\caption{Geographical distribution of the 40 selected countries.   
Color code groups countries into 7 sets: orange (OC) for English speaking countries, 
blue (BC) for former Soviet union ones, 
red (RC) for European ones, 
green (GC) for Latin American ones, 
yellow (YC) for Middle Eastern ones, 
purple (PUC) for North-East Asian ones and 
finally pink (PIC) for South-Eastern countries
(see colors and country names in Table.~\ref{table1} ;
other countries are shown in black).}
\label{fig:monde}
\end{center}
\end{figure}

At present directed networks of real systems can be very large (about  $4.2$ million articles for the English Wikipedia edition in 2013  \cite{eomwiki24} or $3.5$ billion web pages (called also nodes) for a publicly accessible web crawl that was gathered by the Common Crawl Foundation in 2012
\cite{vigna}). For some studies, one might be interested only in the particular interactions between a very small subset of nodes compared to the full network size. For instance, in this paper, we are interested in capturing the interactions of the 40 countries represented in Figure~\ref{fig:monde} using the networks extracted from five Wikipedia language editions covering a few millions of articles each.   
  However, let us assume that there is a rather important person (having 
his own Wikipedia article corresponding to a node $C$) who was
born in country $A$ and worked the main part of his life in country $B$; 
therefore $A$ and $B$ may have links to $C$ (in either direction) and 
thus there may be an indirect link between the two nodes $A$ and $B$ via 
the node $C$ (or other nodes).
In previous works, a solution to this general problem has been proposed in \cite{greduced,politwiki} by defining the reduced Google matrix theory. Main elements of Reduced Google matrix $\GR$ will be presented next, but in a few words, it captures in a 40-by-40 Perron-Frobenius matrix the full contribution of direct and indirect interactions happening in the full Google matrix between the 40 nodes of interest 
(we took top 40 countries of PageRank vector of EnWiki). Elements of reduced matrix $\GR(i,j)$ can be interpreted as the probability for a random surfer starting at webpage $j$ to arrive in webpage $i$ using direct and indirect interactions. Indirect interactions refer to paths composed in part of webpages different from the 40 ones of interest.    
Even more interesting and unique to reduced Google matrix theory, we show here that intermediate computation steps of $\GR$ offer a decomposition of $\GR$ into matrices that clearly distinguish direct from indirect interactions. As such, it is possible to extract the probability for an indirect interaction between two nodes to happen. 

Reduced Google matrix theory is a perfect candidate for analyzing the direct and 
indirect interactions between countries selected worldwide. In this paper, 
we extract from $\GR$ and its decomposition into direct and 
indirect matrices of selected subset network of $N_r=40$ countries. 
The Google matrix of this subset network of $N_r$ 
nodes is computed taking into account  direct and hidden (i.e. indirect) directed links. 
More specifically, we deduce a fine-grained classification of countries that captures 
what we call the \emph{hidden friends} and \emph{hidden followers} of a given country.  
The structure of these graphs provides relevant social information: 
communities of countries with strong ties can be clearly exhibited 
while countries acting as bridges are present as well. 
This is mainly the case for the hidden interactions networks of friends (or followers) 
that offer new information compared to the direct networks of friends (or followers) 
whose topology is mainly enforced by top PageRank countries.     
The mathematical procedure of the reduced $G_R$ matrix construction for $N_r$ nodes
is described in detail in Section 2.

The networks of $\GR$ direct and hidden interactions can be calculated for different Wikipedia language editions. In this paper, reduced Google matrix analysis is applied to the same set of 40 countries on networks representing five different Wikipedia editions: English (EnWiki), Arabic (ArWiki), Russian (RuWiki), French (FrWiki) 
and German (DeWiki) editions. We take for analysis the top 40 countries according to the EnWiki PageRank. 
Wikipedia language editions are usually modified by authors who mainly belong to the region 
associated with this language. Thus our study shows the impact of this cultural bias when comparing direct and hidden networks of friends (or followers) among different language editions. We show that part of the interactions are cross-cultural while others are clearly biased by the culture of the authors. 

\begin{figure*}
\begin{center}
\includegraphics[width=0.95\textwidth]{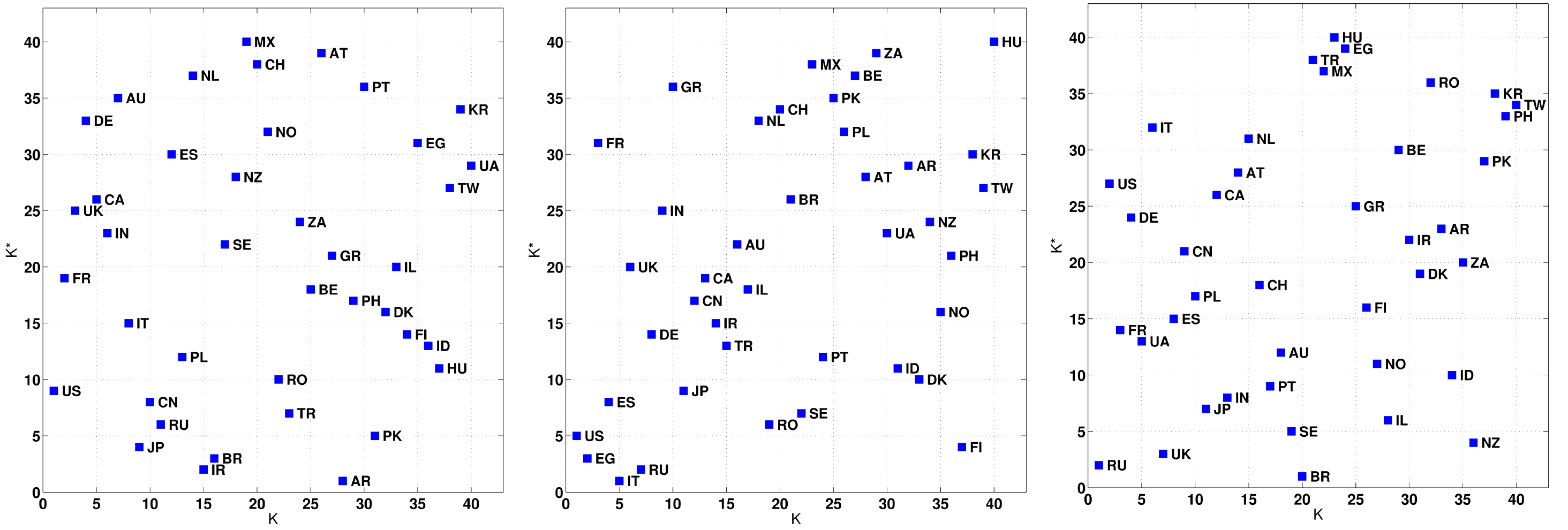}
\caption{Position of countries in the local $(K,K^*)$ plane of the 
reduced network of 40 countries in the EnWiki (left), ArWiki (middle) and RuWiki (right) networks.
}
\label{fig:KK}
\end{center}
\end{figure*}

In  Section~\ref{sec:reduced} we introduce the main elements of reduced Google matrix theory, Section~\ref{sec:matrices} describes $\GR$ calculated for 40 countries and for five different Wikipedia editions. Specific emphasis is given to the very different English, Arabic and Russian editions. Networks of friends and followers for direct and hidden interaction matrices are created and discussed in Section~\ref{sec:networks}, and conclusion is drawn in Section~\ref{sec:conclu}.

\vspace{-0.4cm}
\section{Reduced Google matrix theory}\label{sec:reduced}
\subsection{Google matrix}

It is convenient to describe the network of $N$ Wikipedia articles:
a network node is given by the article name, all nodes are numbered,
some nodes correspond to articles with country names, see Table~1;
the number of countries is much smaller than the total number of articles $N$.
Then the Google matrix $G$is  constructed from the adjacency matrix $A_{ij}$ with elements $1$ if article (node) $j$ points to  article (node) $i$ and zero otherwise. In this case, elements of the Google matrix take the standard form \cite{brin,meyer}
\begin{equation}
  G_{ij} = \alpha S_{ij} + (1-\alpha) / N \;\; ,
\label{eq_gmatrix} 
\end{equation}
where $S$ is the matrix of Markov transitions with elements  $S_{ij}=A_{ij}/k_{out}(j)$, 
$k_{out}(j)=\sum_{i=1}^{N}A_{ij}\neq0$ being the node $j$ out-degree
(number of outgoing links) and with $S_{ij}=1/N$ if $j$ has no outgoing links (dangling node). 
Here $0< \alpha <1$ is the damping factor  which for a random surfer
determines the probability $(1-\alpha)$ to jump to any node;
below we use $\alpha=0.85$.
The right eigenvectors $\psi_i(j)$  of $G$ are defined by:
\begin{equation}
\label{eq_gmatrix2}
\sum_{j'} G_{jj'} \psi_i(j')=\lambda_i \psi_i(j) \; .
\end{equation}
The PageRank eigenvector $P(j)=\psi_{i=0}(j) $ corresponds to the largest 
eigenvalue $\lambda_{i=0}=1$ \cite{brin,meyer}. It has positive elements
which give the probability to find a random surfer on a given node
in the stationary long time limit of the Markov process.
All nodes can be ordered by a monotonically decreasing probability
$P(K_i)$ with the highest probability at $K=1$. The index $K$ 
is the PageRank index.  
The PageRank vector is computed by the PageRank algorithm
with iterative multiplication of initial random vector
by $G$ matrix \cite{brin,meyer,rmp2015}.
Left eigenvectors are biorthogonal to right eigenvectors of different eigenvalues. 
The left eigenvector for $\lambda=1$ has identical (unit) entries  due to the column sum normalization of $G$. 
In the following we use the notations 
$\psi_L^T$  and $\psi_R$ for left and right eigenvectors, respectively. Notation $T$ stands for vector or matrix transposition.

In addition to the matrix $G$ it is useful to introduce a Google matrix $G^*$ constructed from the adjacency matrix of the same network but with inverted direction of all links \cite{linux}. 
The vector $P^*(K^*)$  is called the CheiRank vector \cite{linux,wikizzs} and the index  numbering nodes in order of monotonic decrease of probability $P^*$ is noted as CheiRank index $K^*$. 
Thus, nodes with many ingoing (or outgoing) links have small values of $K=1,2,3...$ (or of $K^*=1,2,3,...$) \cite{meyer,rmp2015}.
We show the distribution of selected countries on the PageRank-ChiRank plane $(K,K^*)$ in Fig.~\ref{fig:KK}.

\subsection{Reduced Google matrix}
We construct the reduced Google matrix for a certain subset of $N_r$ selected nodes
from the global Wikipedia network with $N$ nodes ($N_r \ll N$).
As a subset we choose top 40 countries with the largest  PageRank probabilities 
for EnWiki network (the names are given in Table~1). The reduced Google matrix $G_R$
is constructed on the mathematical basis described below. The main element
of this construction is to keep the same PageRank probabilities of $N_r$ nodes
as in the global network (up to a fixed multiplier coefficient)
and to take into account all indirect links between $N_r$ nodes
coupled by transitions via $N-N_r$ nodes of the global network
(see also \cite{politwiki}).

Let $G$ be a typical Google matrix (\ref{eq_gmatrix}) for 
a network with $N$ nodes such that $G_{ij}\ge 0$ and the 
column sum normalization $\sum_{i=1}^N G_{ij}=1$ is verified. 
We consider a sub-network 
with $N_r<N$ nodes, called ``reduced network''. In this case we can write 
$G$ in a block form:
\begin{equation}
\label{eq_Gblock}
G=\left(\begin{array}{cc}
\Grr & \Grs \\
\Gsr & \Gss \\
\end{array}\right)
\end{equation}
where the index ``$r$'' refers to the nodes of the reduced network and 
``$s$'' to the other $N_s=N-N_r$ nodes which form a complementary 
network which we will call the ``scattering network''. 
Thus $\Grr$ is given by the direct links between selected $N_r$ nodes,
$\Gss$ describes links between all other $N-N_r$ nodes,
$\Grs$ and $\Gsr$ give links between these two parts.
PageRank vector of the full network is given by:
\begin{equation}
\label{eq_Pageank0}
P=\left(\begin{array}{c}
P_r  \\
P_s  \\
\end{array}\right)
\end{equation}
which satisfies the equation $G\,P=P$ 
or in other words $P$ is the right eigenvector of $G$ for the 
unit eigenvalue. This eigenvalue equation reads in block notations:
\begin{eqnarray}
\label{eq_Pagerank1}
({\bf 1}-\Grr)\,P_r-\Grs\,P_s&=&0,\\
\label{eq_Pagerank2}
-\Gsr \,P_r+({\bf 1}-\Gss)\,P_s&=&0.
\end{eqnarray}
Here ${\bf 1}$ is the unit matrix of corresponding size $N_r$
or $N_s$.
Assuming that the matrix ${\bf 1}-\Gss$ is not singular, i.e. all 
eigenvalues $\Gss$ are strictly smaller than unity (in modulus), we obtain 
from (\ref{eq_Pagerank2}) that 
\begin{equation}
\label{eq_Ps}
P_s=({\bf 1}-\Gss)^{-1} \Gsr\,P_r
\end{equation}
which gives together with (\ref{eq_Pagerank1}):
\begin{equation}
\label{eq_Geff1}
\GR P_r=P_r\quad,\quad
\GR=\Grr+\Grs({\bf 1}-\Gss)^{-1} \Gsr
\end{equation}
where the matrix $\GR$ of size $N_r\times N_r$, defined for the 
reduced network, can be viewed as an effective reduced Google matrix. 
Here the contribution of $\Grr$ accounts for direct links 
in the reduced network and the second matrix inverse term 
corresponds to all contributions of indirect links of arbitrary order. 
The matrix elements of $\GR$ are non-negative since the matrix 
inverse in (\ref{eq_Geff1}) can be expanded as:
\begin{equation}
\label{eq_inverse_expand}
({\bf 1}-\Gss)^{-1}=\sum_{l=0}^\infty G_{ss}^{\,l} \;\; .
\end{equation}
In (\ref{eq_inverse_expand}) 
the integer $l$ represents the order of indirect links, i.~e. the number 
of indirect links which are used to connect indirectly two nodes of the 
reduced network. 
We refer the reader to \cite{politwiki} to get the proof that $\GR$ also fulfills the condition 
of column sum normalization being unity.

\begin{figure}[h]
\begin{center}
	\includegraphics[width=0.45\textwidth]{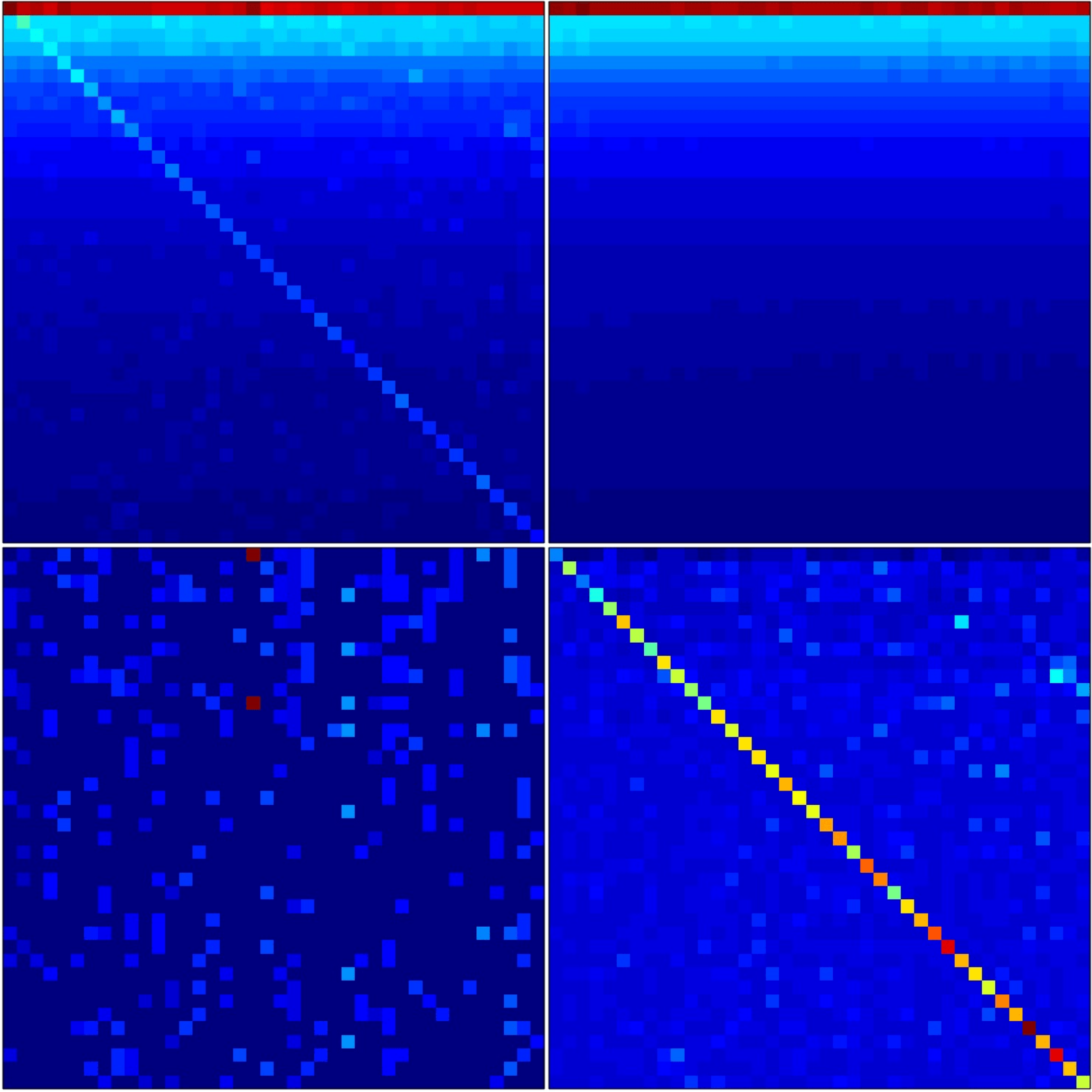}
\caption{Density plots of matrices $\GR$ (top left), $\Gpr$ (top right), $G_{\rm rr}$ (bottom left) and $\Gqr$ (bottom right) for the reduced network of 40 countries in the EnWiki network. The nodes $N_r$ are ordered in lines by increasing PageRank index (left to right) and in columns by increasing PageRank index from top to bottom.  
Color scale represents maximum values in red ($0.15$ in top panels; $0.01$ in bottom left panel;
$0.03$ in bottom right panel), intermediate in green and minimum (approximately zero) in blue. 
}
\label{fig:EnWiki}
\end{center}
\end{figure}

\subsection{Numerical evaluation of $\GR$}
We can question how to evaluate practically the expression 
(\ref{eq_Geff1}) of $\GR$ for a particular sparse and quite large 
network when $N_r\sim 10^2$-$10^3$ is small compared to $N$ and $N_s \approx N\gg N_r$. 
If $N_s$ is too large (e.~g. $N_s > 10^5$) a direct naive evaluation 
of the matrix inverse $({\bf 1}- \Gss)^{-1}$ in (\ref{eq_Geff1}) 
by Gauss algorithm is not efficient. In this case we can try the 
expansion (\ref{eq_inverse_expand}) provided it converges sufficiently 
fast with a modest number of terms. However, this is most likely not the 
case for typical applications since $\Gss$ is very likely to have 
at least one eigenvalue very close to unity. 

Therefore, we consider the situation 
where the full Google matrix has a well defined gap between the leading 
unit eigenvalue and the second largest eigenvalue (in modulus). For example 
if $G$ is defined using a damping factor $\alpha$ in the standard way, 
as in (\ref{eq_gmatrix}), the 
gap is at least $1-\alpha$ which is $0.15$ for the standard choice 
$\alpha=0.85$ \cite{meyer}.
In order to evaluate the expansion 
(\ref{eq_inverse_expand}) efficiently, we need to take out analytically 
the contribution of the leading eigenvalue of $\Gss$ close to unity which is 
responsible for the slow convergence.

Below we denote by $\lambda_c$ this leading eigenvalue of $\Gss$ and by $\psi_R$ 
($\psi_L^T$) the corresponding right (left) eigenvector such that 
$\Gss\psi_R=\lambda_c\psi_R$ (or $\psi_L^T \Gss=\lambda_c\psi_L^T$). 
Both left and right eigenvectors as well as $\lambda_c$ can be efficiently 
computed by the power iteration method in a similar way as the standard 
PageRank method. 
Vectors $\psi_R$ are normalized with $E_s^T\psi_R=1$ and $\psi_L$ with $\psi_L^T\psi_R=1$. 
It is well known (and easy to show) that $\psi_L^T$ is orthogonal to all other 
right eigenvectors (and $\psi_R$ is orthogonal to all other 
left eigenvectors) of $\Gss$ with eigenvalues different from $\lambda_c$. 
We introduce the operator ${\cal P}_c=\psi_R\psi_L^T$ which is the 
projector onto the eigenspace of $\lambda_c$ and we denote by 
${\cal Q}_c={\bf 1}-{\cal P}_c$ the complementary projector. 
One verifies directly that both projectors commute with the matrix $\Gss$ 
and in particular ${\cal P}_c \Gss=\Gss{\cal P}_c=\lambda_c{\cal P}_c$. 
Therefore we can derive:
\begin{eqnarray}
\label{eq_inverse_project1}
({\bf 1}-\Gss)^{-1}&=
&{\cal P}_c\frac{1}{1-\lambda_c}+
{\cal Q}_c \sum_{l=0}^\infty \bar G_{ss}^{\,l}
\end{eqnarray}
with $\bar G_{ss}={\cal Q}_c \Gss{\cal Q}_c$ and using the 
standard identity ${\cal P}_c{\cal Q}_c=0$ for complementary 
projectors. 
The expansion in (\ref{eq_inverse_project1}) converges rapidly since $\bar G_{ss}^{\,l}\sim |\lambda_{c,2}|^l$ 
 with $\lambda_{c,2}$ being the second largest eigenvalue which is significantly lower than unity.

The combination of (\ref{eq_Geff1}) and (\ref{eq_inverse_project1}) 
provides an explicit algorithm feasible for a numerical implementation 
for modest values of $N_r$, large values of $N_s$ and of course 
if sparse matrices $G$, $\Gss$ are considered.  
We refer the reader to \cite{politwiki} for more advanced implementation considerations.
\begin{figure*}[]
\begin{center}
	\includegraphics[width=0.95\textwidth]{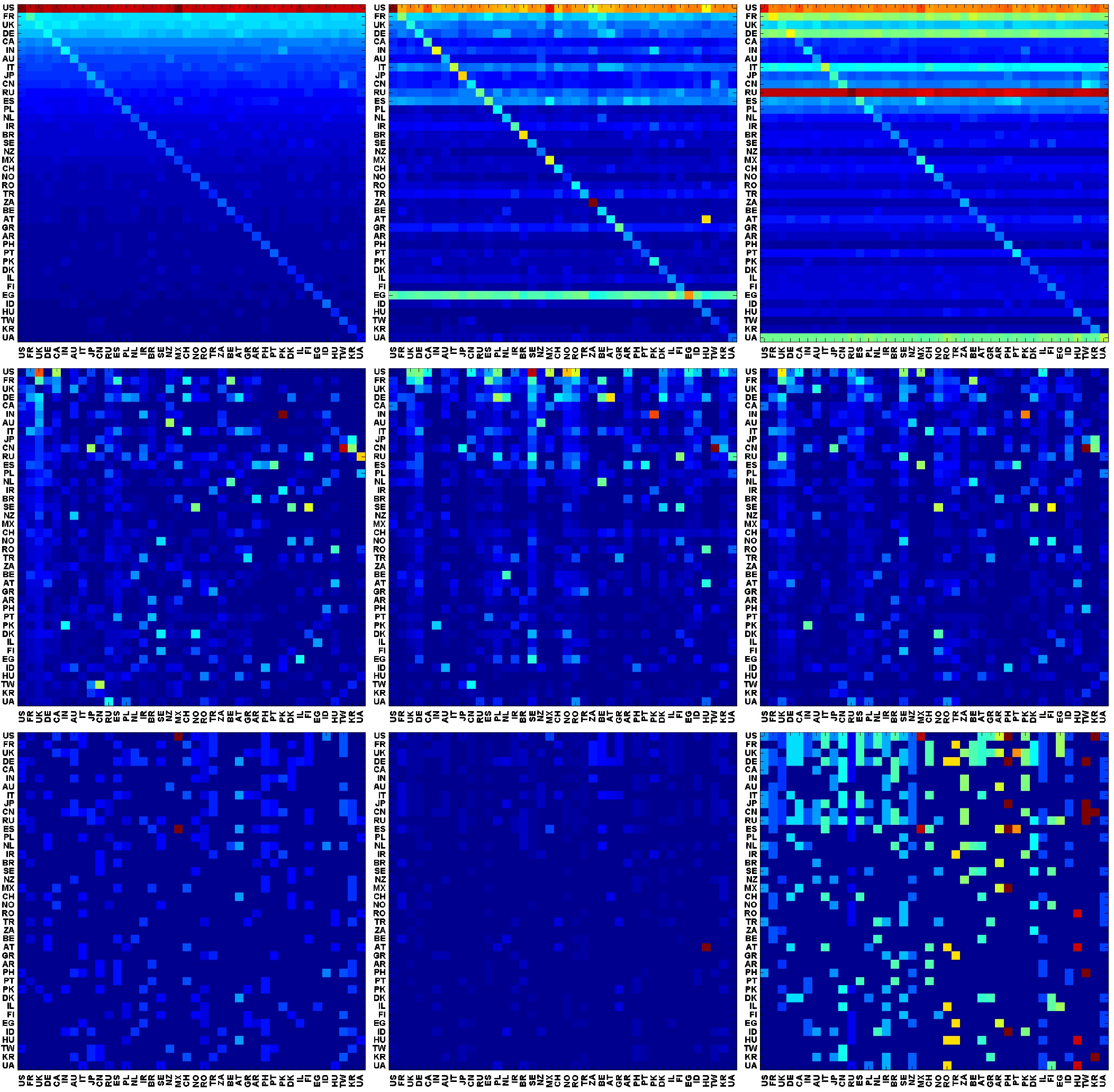}
\caption{Density plots of the matrix elements of  $\GR$ (first line), $\Gqrnd$ (second line) and 
$\Grr$ (third line) for the reduced network of 40 countries of EnWiki (left column), 
ArWiki (middle column) and RuWiki (right column). The country names are given 
on the axes in the order names in Table 1, thus
the nodes $N_r$ are ordered in lines and columns by the reference PageRank of EnWiki.
The colors represent maximum (red corresponds to:  0.15, 0.19, 0.13 in top panels from left to right;
0.01, 0.03, 0.012 in middle panels and
0.01, 0.011, 0.006 in bottom panels respectively), intermediate (green) and minimum (blue for zero) 
values for a give matrix.
}
\label{fig:density}
\end{center}
\end{figure*}

\begin{figure*}[htb]
\begin{center}
	\includegraphics[width=0.95\textwidth]{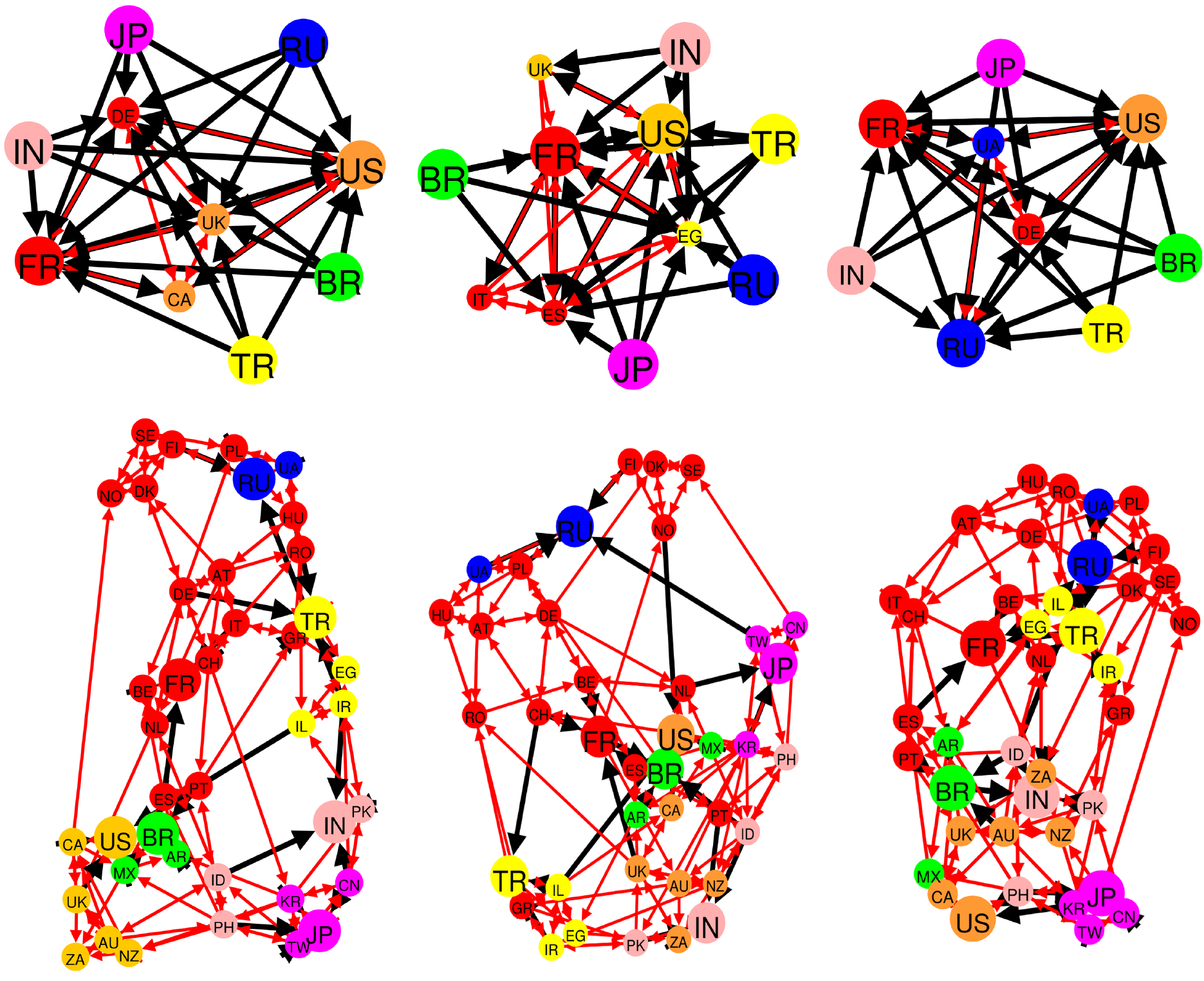}
\caption{Network structure of friends (top line) and followers (bottom line) induced by the 7 top countries of each geographical area (US, FR, IN, JP, BR, TR, RU) in $\GR$. 
Results are plotted for EnWiki (left column), ArWiki (middle column) and RuWiki (right column). Node colors represent geographic appartenance to a group of countries (cf. Table~\ref{table1} for details).
Top (bottom) graphs: the top country node points (is pointed by) with a bold black arrow to its top 4 friends (followers). Red arrows show friends of friends (resp. followers of followers) interactions computed until no new edges are added to the graph. Drawn with \cite{graphpack}.
}
\label{fig:networkGR}
\end{center}
\end{figure*}
\begin{figure*}[htb]
\begin{center}
	\includegraphics[width=0.95\textwidth]{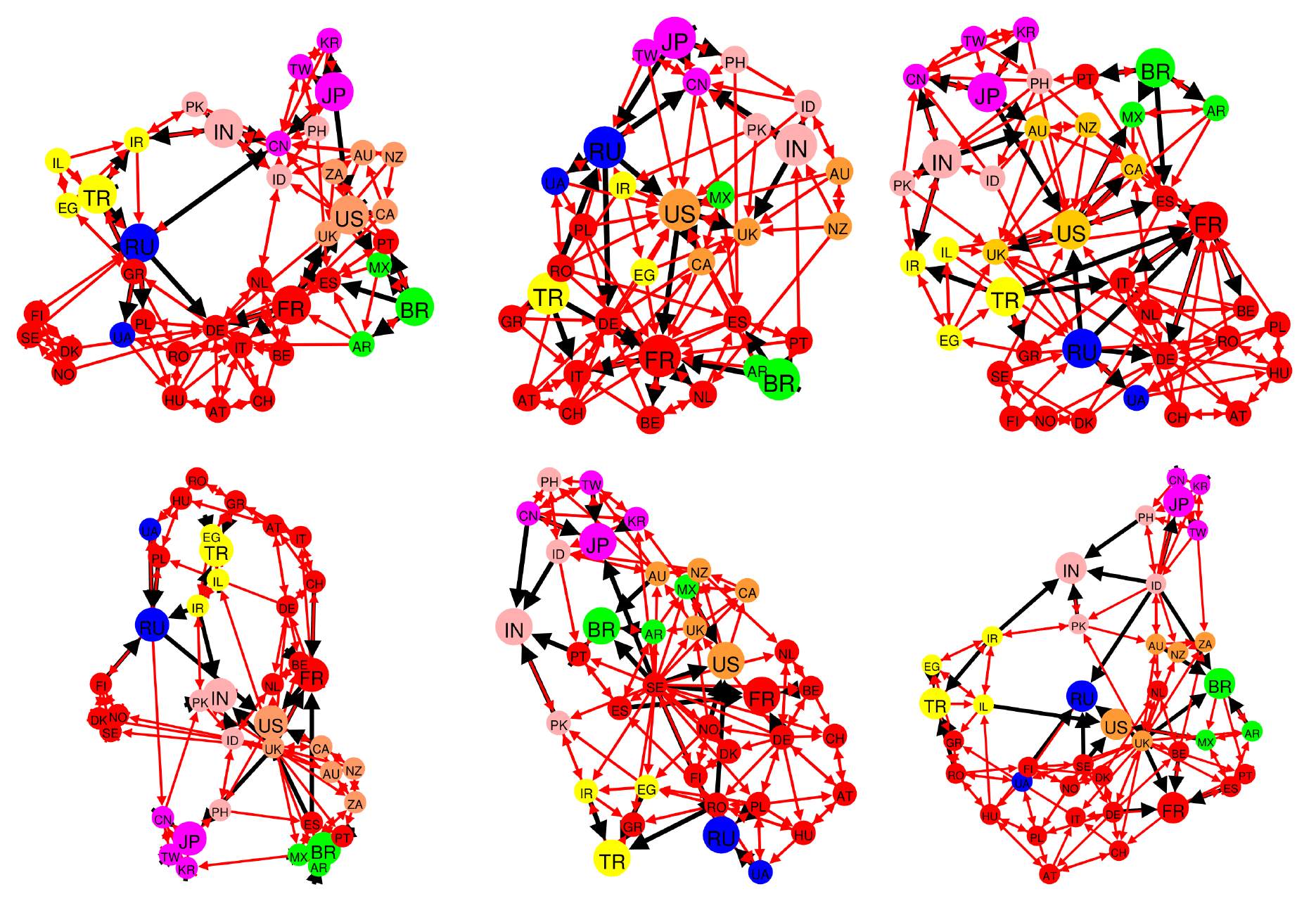}
\caption{Same legend as Fig.~\ref{fig:networkGR} except friends and followers are computed from $\Gqrnd$ (hidden relationships).
}
\label{fig:networkGqr}
\end{center}
\end{figure*}

\subsection{Decomposition of $\GR$}

On the basis of equations 
(\ref{eq_Geff1})-(\ref{eq_inverse_project1}),
the reduced Google matrix can be presented as a sum of three components:
\begin{equation}
\label{eq_3terms}
\GR=\Grr + \Gpr + \Gqr ,
\end{equation}
with the first component $\Grr$ given by direct matrix elements of $G$
among the selected $N_r$ nodes.
The second projector component $\Gpr$ is given by:
\begin{equation}
\label{eq_2ndterm}
\Gpr =  \Grs  {\cal P}_c \Gsr/(1-\lambda_c) , \; 
{\cal P}_c=\psi_R\psi_L^T \;.
\end{equation}

The third component $\Gqr$ is of particular interest in this study as it characterizes the impact of indirect or hidden links. It is given by:
\begin{eqnarray}
\label{eq_3rdterm}
\nonumber
\Gqr &=&  \Grs [{\cal Q}_c \sum_{l=0}^\infty \bar G_{ss}^{\,l}]  \Gsr , \; 
{\cal Q}_c={\bf 1}-{\cal P}_c, \;\\
\bar G_{ss} &=& {\cal Q}_c \Gss{\cal Q}_c .
\end{eqnarray}

We do characterize the strength of these 3 components by their respective weights $\Wrr$, $\Wpr$, $\Wqr$ given respectively by the sum of all matrix elements of  $\Grr$, $\Gpr$, $\Gqr$ divided by $N_r$. By definition we have $\Wrr + \Wpr + \Wqr =1$.

Reduced Google matrix has been computed, together with its components $\Grr$, $\Gpr$ and $\Gqr$, for the English language edition of Wikipedia (EnWiki) and for the $N_R=40$ countries listed in Table~\ref{table1}. These countries are the ones with top PageRank $K$ in the network of EnWiki. Density plots of $\GR$, $\Grr$, $\Gpr$ and $\Gqr$, are given in Figure \ref{fig:EnWiki}. Countries are ordered by increasing $K$ value.  The weight of the three matrix components of $\GR$ are $\Wpr=0.96120$, $\Wqr= 0.029702$ and $\Wrr = 0.009098$. Predominant component is clearly $\Gpr$ but as we will explain next, it is not the most meaningful.

The meaning of $\Grr$ is clear as it is directly extracted from the global Google matrix $G$. It gives the direct links between the selected nodes and more specifically the probability $\Grr(i,j)$ for the surfer to go directly from column $j$ country to line $i$ country. 
However, since each column is normalized by the number of outgoing links, absolute probabilities cannot be compared to each other across columns.   

The sum of $\Gpr$ and $\Gqr$ represents the contribution of all indirect links through the scattering matrix $\Gss$. As seen on Figure.~\ref{fig:EnWiki}, the projector component $\Gpr$ is composed of nearly identical columns. Moreover, values of each column are proportional to the PageRank of the countries (lines and columns are ordered by increasing $K$ values). As detailed in \cite{politwiki}, we observe numerically that $\Gpr \approx P_r\,E_r^T/(1-\lambda_c)$, meaning that each column is close to the normalized vector $P_r/(1-\lambda_c)$. As such, $\Gpr$ transposes essentially in $\GR$ the contribution of the first eigenvector of $G$. We can conclude that even if the overall column sums of $\Gpr$ account for $\sim 95$-97\% of the total column sum of $\GR$, $\Gpr$ doesn't offer innovative information compared to PageRank analysis. 

A way more interesting contribution is the one of $\Gqr$. This matrix captures higher-order indirect links between the $N_r$ nodes due to their interactions with the global network environment. We will refer to these links as \emph{hidden links}. 
We note that $\Gqr$ is composed of two parts $\Gqr = \Gqrd + \Gqrnd$ where the first term gives only the diagonal part of the matrix $\Gqrd$ and thus represents the probabilities to stay on the same node during multiple iterations of $\bar G_{ss}$ in (\ref{eq_3rdterm}) while the second matrix captures only non-diagonal terms in $\Gqrnd$. As such, $\Gqrnd$ represents indirect (hidden) links between the $N_r$ nodes appearing via the global network.
We note that certain matrix elements of $\Gqr$ can be negative, which is possible due to the negative terms in ${\cal Q}_c={\bf 1}-{\cal P}_c$ appearing in (\ref{eq_3rdterm}). The total weight of negative elements is however much smaller than $\Wqr$ (at least 6 times smaller and even non-existing in ArWiki). 
Of course, the full reduced Google matrix $G_R$ has only positive or zero matrix elements.
In the following, our study concentrates mainly on the meaning of $\Gqrnd$, emphasizing the meaning of its largest positive values in the study of Section~\ref{sec:networks}. 

\vspace{-0.3cm}
\section{Matrices of world countries}\label{sec:matrices}
Our study focuses on the networks representing 5 different Wikipedia editions\footnote{Data collected mid February 2013.}  from the set of 24 analyzed in \cite{eomwiki24}: EnWiki, ArWiki, RuWiki, DeWiki and FrWiki that contain 4.212, 0.203 , 0.966, 1.533 and 1.353 millions of articles each.

\subsection{Selected countries}
The 40 countries listed in Table~\ref{table1} have been selected from the EnWiki network after computing the PageRank for the complete network. The 40 countries with largest PageRank probability have been chosen, and ordered by a local PageRank index $K$
varying between $1$ and $40$. The most influential countries are found to be at the top values $K=1,2,...$. In addition we determine the local CheiRank index $K^*$ of the selected countries using the CheiRank vector of the global network \cite{linux,rmp2015}. At the top of $K^*$ we have the most communicative countries. Table~\ref{table1} lists $K$ and $K^*$ for EnWiki, ArWiki and RuWiki. Not surprisingly, the order of top countries changes with respect to the edition (for instance, the top country for $K$ is US except for RuWiki whose top country is Russia).

Countries that belong to the same region or having a common piece of history may probably exhibit stronger interactions in Wikipedia. As such, we have created a color code that groups together countries that either belong to the same geographical region (e.g. Europe, Latin America, Middle East, North-East Asia, South-East Asia) or share a big part of history (former USSR; English speaking countries that are the legacy of the former British Empire). Color code can be seen either in Figure~\ref{fig:monde} or in the text color of Table~\ref{table1}. 

It is convenient as well to plot all nodes in the ($K$, $K^*$) plane to highlight the countries that are the most influential ($K=1, 2,$ ...) and the most communicative ($K^*=1, 2,$ ...) at the same time. Figure~\ref{fig:KK} plots all 40 countries in the ($K$, $K^*$) plane for EnWiki, ArWiki and RuWiki editions. This plot is a bi-objective plot where $K$ and $K^*$ are to be minimized concurrently. It is interesting to look at the set of non-dominated countries which are the ones such that there is no other country beating them for both $K$ and $K^*$. Cultural bias is obvious here as for EnWiki, this set is composed of \{US, JP, IR, AR\}, for ArWiki of \{US, EG, IT\} and for RuWiki of only Russia. 
We note that according to Figure~\ref{fig:KK} 
some countries with high $K$ value 
(relative few in-degree and low PageRank probability) 
act as important  diffusers of content (low $K^*$)
even for the language editions being 
different from the language spoken in those countries 
(BR, AR, RO).
Also countries with low $K$ (having many citations) 
are relatively poor diffusers (FR, CA).
Thus  Figure~\ref{fig:KK} demonstrates 
that different cultures attribute different 
degree of country popularity  (PageRank probability and $K$ index)
and different communicative degree (CheiRank probability and $K^*$ index).
This indicates the nontrivial features of cultures propagation and interactions.

\subsection{Density plots of $\GR$, $\Grr$ and $\Gqrnd$}

For the three EnWiki, ArWiki and RuWiki editions, Figure~\ref{fig:density} plots the density of matrices $\GR$, $\Gqrnd$ and $\Grr$. We  keep for all plots the same order of countries extracted from the EnWiki network. This is meant to highlight cultural differences among Wikipedia editions. In the $\GR$ plots, it is clear that this matrix is dominated by the projector $\Gpr$ contribution, which is proportional to the global PageRank probabilities. In $\GR$, not surprisingly, the cultural bias is pronounced due to its strong tie to PageRank. For instance, Egypt is much more important in ArWiki than in other editions, and Russia is the top country in RuWiki.
 
The information from direct links between countries is provided by $\Grr$. As expected, the per-column normalization prevents a meaningful per-line analysis. For EnWiki and ArWiki, the respective columns of Mexico and Hungary are predominant due to their little number of outgoing links. On the contrary, $\Gqrnd$ offers a much more unified view of countries interactions as it seems to highlight more general interaction that are less biased by cultural views. 
For instance, for these three Wikipedia editions, the hidden links connecting Taiwan to China and Pakistan to India are really strong in $\Gqrnd$.  The link connecting Ukraine to Russia is very strong in EnWiki and ArWiki. It is surprisingly absent from RuWiki (or maybe this is exactly a cultural bias we are observing since during a long time both countries were part of USSR, and there was thus no specific difference between them). 
Other interesting hidden links are highlighted in EnWiki as New-Zealand is directed to Australia or in RuWiki linking Canada to the USA.

\subsection{Friends and followers}

In order to better capture the interactions provided by $\Gqrnd$ and $\GR$, we have listed for all 5 Wikipedia editions the top 4 friends and top 4 followers of a set of 7 leading countries. One leading country per group has been selected: US for English speaking countries, Brazil for Latin America,  France for Europe, Japan for North-East Asia, India for South-East Asia, Russia for the Soviet block and Turkey for the Middle-East. To pick them inside each group, we have chosen the country whose worst PageRank order over all 5 Wikipedia editions is the highest. 

For each leading country, we extract from both matrices $\Gqrnd$ and $\GR$ the top 4 \emph{Friends} (resp. \emph{Followers}) of country $j$ given by the 4 best values of the elements of column $j$ (resp. of line $j$). In other words, it corresponds to destinations of the 4 strongest outgoing links of $j$ and the countries at the origin of strongest 4 ingoing links. A summary of relevant results is given in Table~\ref{table2} to show cross-culture interactions. 

Looking at $\GR$ friends, top friends of leading countries are strongly related to the top PageRank countries (we have a predominance of US, France and Germany). Similarly, cross-edition followers of US are Mexico and Canada, and followers of Japan are China, Korea and Taiwan. On the opposite, higher-order interactions of $\Gqrnd$ are not as much influenced by PageRank. More subtle but realistic interactions appear: Canada is always identified as a hidden friend of USA while it was never the case in $\GR$. Similarly, Ukraine is always tagged as a hidden friend of Russia; Italy and Spain as friends of France. Thus $\Gqrnd$ seems to emphasize more fine-grained regional interactions. Next section exploits the concept of friends and followers to create new network representations derived from $\GR$ and $\Gqrnd$. 

\section{Networks of 40 countries}\label{sec:networks}

This study concentrates again on the same 7 leading countries as before. Top 4 friends and top 4 followers of these leading countries are extracted from $\GR$ and $\Gqrnd$ to plot the graphs of Figures~\ref{fig:networkGR} and \ref{fig:networkGqr}, respectively. Note that Figure~\ref{fig:networkGqr} essentially highlights hidden links. The black thick arrows identify the top 4 friends and top 4 followers interactions. Red arrows represent the friends of friends (respectively the followers of followers) interactions that are computed recursively until no new edge is added to the graph. All graphs are plotted using a force direct layout. 

The networks of friends obtained from $\GR$ never expand to the full set of 40 nodes. They only concentrate on about 10 countries (including the 7 leading ones). This happens since $\GR$ is dominated by the projector component. Looking at the follower graphs, more information can be observed. North-East Asian, Middle-Eastern and Latin American create, in all editions, a cluster of nodes densely interconnected. European countries enclose Russia and Ukraine as these countries are linked to EU countries that were part of the former Soviet Union zone of influence (e.g. Romania, Hungary, Finland, etc.). The networks of followers end up almost spanning the full set of 40 countries.  

The networks of friends obtained from $\Gqrnd$ don't concentrate to a limited set of countries as it is the case for $\GR$. They end up spanning the full set of countries. The hidden friend links show that the interactions between the geographical groups are coherent.  North-East Asian countries are linked to South-East Asian countries and to English speaking countries in EnWiki and RuWiki.  Interestingly,  the set of Baltic countries (SE, NO, DK, FI) create most of the time full meshes, and interconnect Europe and Russia. 
Cultural bias can be observed as well in these plots. For non-Arabic editions, Middle-Eastern countries create a well-connected cluster of nodes. But for ArWiki,  Turkey exhibits a stronger connection with Europe than with the other Middle-Eastern countries. In the view of Arabic countries, Turkey is seen closer to Europe than others for sure. 

\section{Conclusion}  
\label{sec:conclu}
This work offers a new perspective for future geopolitics studies. 
It is possible to extract from multi-cultural Wikipedia networks 
a global understanding of the interactions between countries 
at a global, continental or regional scale. 
On the basis of the reduced Google matrix analysis of the Wikipedia network
we determined the relations of friends and followers 
of countries on purely statistical mathematical grounds 
obtained from the huge knowledge accumulated by
Wikipedia editions in different languages.
This approach provides results independent of cultural bias.
The reduced Google matrix theory has been shown to capture 
hidden and indirect interactions among countries, 
resulting in new knowledge on geopolitics.   

\vspace{-0.2cm}
\section*{Acknowledgements}
This work was supported by APR 2015 call of University of Toulouse and by R\'egion Occitanie (project GOMOBILE), MASTODONS-2016 CNRS project APLIGOOGLE, and EU CHIST-ERA MACACO project ANR-13-CHR2-0002-06.

\begin{table}[h]
\begin{center}
{\relsize{-2}
\vspace{-0.3cm}
\include{table1}
}
\caption{List of names of 40 selected countries with PageRank $K$, 
CheiRank $K^*$ for EnWiki, ArWiki and RuWiki, ordered by increasing PageRank of EnWiki edition. 
Figure~\ref{fig:monde} gives color correspondence details.
A Wikipedia article 
with country name represents 
one node of the whole network with $N$ nodes,
e.g. https://en.wikipedia.org/wiki/France for ``France'' in EnWiki.
}
\label{table1}
\end{center}
\end{table}

\newpage
\begin{table*}
\begin{center}
{\relsize{-2}
\include{table2}
\vspace{-0.2cm}
\include{table3}
}
\caption{Cross-edition direct friends and followers extracted from $\GR$ matrix (top table) and from $\Gqrnd$ (bottom table) for the top countries of each area. For each top country, we list the direct friends (followers) present in the direct friends list given by all five Wikipedia editions, the ones present in 4 editions out of 5 and the ones present in 3 editions out of five. }
\label{table2}
\end{center}
\end{table*}


\end{document}